\documentclass[conference]{IEEEtran}

\usepackage{cite}
\usepackage{amsmath,amssymb,amsfonts}
\usepackage{graphicx}
\usepackage{textcomp}
\usepackage{xcolor}
\usepackage{tikz}
\usepackage{enumitem}
\usepackage{listings}
\usepackage{xcolor}
\usepackage{booktabs}
\usepackage{algpseudocode}
\usepackage{float}
\usepackage{adjustbox}
\usepackage{multirow}
\usepackage{pgfplots}
\usepackage{caption}
\usepackage{subcaption}
\usepackage{colortbl} 
\usepackage{balance}
\usepackage{xspace}
\usepackage{soul}
\usepackage{algorithm}

\setlist[enumerate]{leftmargin=0.9cm}

\algrenewcommand\algorithmicrequire{\textbf{Input:}}
\algrenewcommand\algorithmicensure{\textbf{Output:}}

\definecolor{codegreen}{rgb}{0,0.6,0}
\definecolor{codegray}{rgb}{0.5,0.5,0.5}
\definecolor{codepurple}{rgb}{0.58,0,0.82}
\definecolor{backcolour}{rgb}{0.95,0.95,0.92}

\definecolor{susanColor}{RGB}{255, 204, 204} 
\definecolor{lqiColor}{RGB}{204, 229, 255} 
\definecolor{ssColor}{RGB}{204, 255, 204} 
\definecolor{fftColor}{RGB}{255, 255, 204} 

\definecolor{shellbg}{gray}{0.95} 
\definecolor{shelltext}{rgb}{0,0,0}
\definecolor{shellstring}{rgb}{0.0,0.0,0.0}

\lstdefinelanguage{shell}{
    sensitive=false, 
    morecomment=[l]{\$}, 
    morecomment=[s]{/*}{*/}, 
    morestring=[b]" 
} %
\lstdefinestyle{shellst}{  
    backgroundcolor=\color{shellbg},   
    commentstyle=\color{shelltext}\bfseries,
    numberstyle=\tiny\color{codegray},
    stringstyle=\color{shellstring},
    basicstyle=\ttfamily,      
    frame=single,
    breaklines=true,                 
    captionpos=b,                    
    keepspaces=true,                 
    numbers=left,                    
    numbersep=10pt,                  
    showspaces=false,                
    showstringspaces=false,
    showtabs=false,
    columns=flexible,
}

\lstdefinestyle{codest}{
    backgroundcolor=\color{shellbg},   
    commentstyle=\itshape\color{purple!40!black},
    keywordstyle=\bfseries\color{green!40!black},
    numberstyle=\tiny\color{codegray},
    stringstyle=\color{magenta},
    basicstyle=\ttfamily\small,
    identifierstyle=\color{blue},
    breakatwhitespace=false,
    breaklines=true,                 
    captionpos=b,                    
    keepspaces=true,                 
    numbers=left,                    
    numbersep=5pt,                  
    showspaces=false,                
    showstringspaces=false,
    showtabs=false,                  
    tabsize=2
}

\newcommand{\tool}{Approxify\xspace}
\newcommand{\firstcomp}{Code Parser\xspace}
\newcommand{\secondcomp}{ApproxSafe Mapper\xspace}
\newcommand{\thirdcomp}{Output Validator\xspace}
\newcommand{\fourthcomp}{Approximator\xspace}

\pgfplotsset{compat=1.18}

\def\BibTeX{{\rm B\kern-.05em{\sc i\kern-.025em b}\kern-.08em
    T\kern-.1667em\lower.7ex\hbox{E}\kern-.125emX}}
\begin{document}

\title{Approxify: Automating Energy-Accuracy Trade-offs in Batteryless IoT Devices}

\author{\IEEEauthorblockN{Muhammad Abdullah Soomro}
\IEEEauthorblockA{\textit{LUMS} \\ Pakistan}
\and
\IEEEauthorblockN{Naveed Anwar Bhatti}
\IEEEauthorblockA{\textit{LUMS} \\ Pakistan}
\and
\IEEEauthorblockN{Muhammad Hamad Alizai}
\IEEEauthorblockA{\textit{LUMS} \\ Pakistan}}

\maketitle

\begin{abstract}
Batteryless IoT devices, powered by energy harvesting, face significant challenges in maintaining operational efficiency and reliability due to intermittent power availability. Traditional checkpointing mechanisms, while essential for preserving computational state, introduce considerable energy and time overheads. This paper introduces \tool, an automated framework that significantly enhances the sustainability and performance of batteryless IoT networks by reducing energy consumption by approximately 40\% through intelligent approximation techniques. \tool balances energy efficiency with computational accuracy, ensuring reliable operation without compromising essential functionalities.
Our evaluation of applications, SUSAN and Link Quality Indicator (LQI), demonstrates significant reductions in checkpoint frequency and energy usage while maintaining acceptable error bounds.
\end{abstract}

\begin{IEEEkeywords}
Sustainable IoT, Approximate Computing, Energy-Efficient Networking, Low Power Wireless Networks
\end{IEEEkeywords}

\section{Introduction}
\label{sec:intro}

The rise of batteryless IoT devices, powered by energy harvesting, introduces significant challenges for computation and communication, highlighting the importance of energy efficiency and continuous operation for IoT applications~\cite{lucia2017intermittent,ahmed2021survey}. Traditional checkpointing methods, while useful for coping with power interruptions by saving and restoring system states, impose considerable overheads~\cite{bhatti2017harvos,alharbi2023checkpointing}. Although checkpointing helps maintain computational progress, it significantly consumes time and energy, detracting from the energy available for core tasks and thus negatively impacting the system's efficiency and output. This challenge is critical in the context of energy-efficient and green networking, where minimizing energy consumption is paramount to ensuring the scalability and reliability of IoT deployments.

Approximate computing emerges as a strategic complement to checkpointing, balancing computational accuracy and energy efficiency, as highlighted in recent studies~\cite{javed2023moptic,bambusi2022case,ganesan2019s,10.1145/3581791.3596845,9923863, 10.1145/3607918, lin2023intermittent}.  It enables systems to perform tasks more energy-efficiently under intermittent power conditions, which is especially relevant for \emph{IoT applications where slight inaccuracies are tolerable}. Given the fluctuating energy sources typical of these systems, approximation aligns with the primary goal to maximize energy utilization while still producing effective computational outcomes, fundamentally aiming to reduce the number of checkpoints (power cycles) required to complete applications. The reduction in checkpoints could potentially lead to improved network performance, such as lower latency and increased throughput in data processing tasks.


A recurring theme in recent studies~\cite{javed2023moptic,bambusi2022case,ganesan2019s,10.1145/3581791.3596845,9923863, 10.1145/3607918, lin2023intermittent} is the tailored nature of each implementation of approximation techniques, requiring application-specific fine-tuning. However, these specialized implementations often lack generalizability to different applications or even the same application under different conditions, highlighting the need for a more flexible and automated approach to applying approximate computing across diverse batteryless IoT environments. Such a method would enable broader applications of approximate computing in batteryless IoT settings.

Nonetheless, applying approximation techniques in such settings is challenging for the following reasons:

\begin{enumerate}[label=\textbf{[C\arabic{enumi}]}]
\item \textbf{Complex manual tuning:} Approximation techniques necessitate detailed, manual adjustment for each application, rendering the process time-consuming and susceptible to inaccuracies. Balancing energy efficiency with accuracy is essential for optimal performance.

\item \textbf{Identifying approximation opportunities:}  Implementing approximations requires a deep understanding of both the application and techniques, as these decisions greatly affect energy efficiency and output reliability, making them crucial for leveraging approximate computing.

\item \textbf{Impact assessment and behavior prediction:} Evaluating the effects of approximations and forecasting program behavior after deployment poses considerable
challenges, thereby complicating the scalability and adaptability of solutions across diverse use cases.
\end{enumerate}

To address these challenges, we introduce \tool, a tool designed to automate the application of approximation techniques. \tool features an intelligent, user-friendly design that minimizes developer effort. Different components of \tool work in tandem to analyze code, select blocks for approximation, and employ strategies to enhance energy efficiency without sacrificing accuracy. \tool stands apart from existing approximation frameworks~\cite{baek2010green, samadi2013sage, samadi2014paraprox} as it aims to reduce checkpoint frequency rather than merely minimizing error, a critical adjustment for batteryless IoT devices. Additionally, \tool leverages user-supplied data traces to refine approximation parameters, ensuring optimizations match real-world scenarios and boosting both accuracy and efficiency.

We benchmarked \tool across two IoT applications in different energy harvesting scenarios, confirming its efficacy in autonomously implementing approximations, enhancing energy efficiency, and maintaining acceptable output fidelity. These results suggest that \tool can significantly aid users in adopting approximation techniques in batteryless IoT applications, optimizing performance in energy-constrained environments.

\section{\tool}

\subsection{Design Goals}
Our design goals for \tool address the key challenges in applying approximate computing to batteryless embedded systems, as discussed in Section~\ref{sec:intro}. These goals shape \tool's architecture to facilitate effective approximations:

\begin{enumerate}[label=\textbf{[D\arabic{enumi}]}]
     \item \textbf{Broad application support:} Versatile across applications, countering the limitation of approximation techniques to specific use-cases \textbf{[C1]}. A robust parsing system generalizes approximation opportunities, enhancing developer flexibility.

    \item \textbf{Minimized developer footprint:} Simplifies the developer's experience \textbf{[C1]} by automating the integration of approximation strategies, allowing developers to focus on core functionality.

    \item \textbf{Comprehensive approximation techniques:} Includes a range of methods to balance energy efficiency and accuracy, addressing \textbf{[C2]} and enabling informed decisions on the best strategies for each application.

    \item \textbf{Dynamic approximation calibration:} Evaluates and adjusts approximation levels at compile time by simulating the application under realistic energy traces. This tackles challenges related to manual tuning and behavior prediction, addressing \textbf{[C1]} and \textbf{[C3]}.
\end{enumerate}


\subsection{Approximation Techniques}
\tool utilizes software-based approximation techniques, avoiding the need for specialized hardware, which is crucial for the targeted platforms that typically lack built-in hardware approximations. The focus of \tool is not on creating new approximation methods but on automating the identification of appropriate approximation opportunities within applications. \tool employs the following techniques:

\subsubsection{\textbf{Loop Perforation}} Loop perforation sacrifices accuracy for performance by executing only a subset of loop iterations. \tool uses three techniques:

\noindent
\textbf{Truncation Perforation:} Early loop termination using a tunable \texttt{truncation factor} (0.00 to 1.00), which skips iterations from the loop's end.
          
\noindent
\textbf{Sampling Perforation}: Adjusts the loop's iteration rate by executing every $i$-th iteration, where $i$ is the \texttt{sampling factor}. For example, a sampling factor of 2 executes every second iteration, skipping 50\% of iterations, while a factor of 3 executes every third iteration, skipping approximately 66.7\%. This method balances performance and computational efficiency by reducing the number of iterations proportionally.

\noindent
\textbf{Random Perforation}: Uses a randomly generated number to decide whether to skip each loop iteration. This method introduces minimal computational overhead and is particularly effective for loops where each iteration is critical, and other perforation methods might lead to unacceptable errors. \tool uses a \texttt{randomization threshold} to fine-tune the balance between accuracy and performance, allowing for flexible adjustment of the approximation level.

\subsubsection{\textbf{Function Approximations}} \tool enhances traditional memoization with a tolerance-based approach, allowing cached results to be reused for inputs similar within a predefined tolerance range. This reduces the frequency of function executions, balancing energy efficiency and accuracy. The \texttt{tolerance range} is carefully chosen to ensure that the approximation maintains a balance between energy efficiency and the accuracy requirements of the application.


\begin{figure}
    \centering
    \includegraphics[width=2.7in]{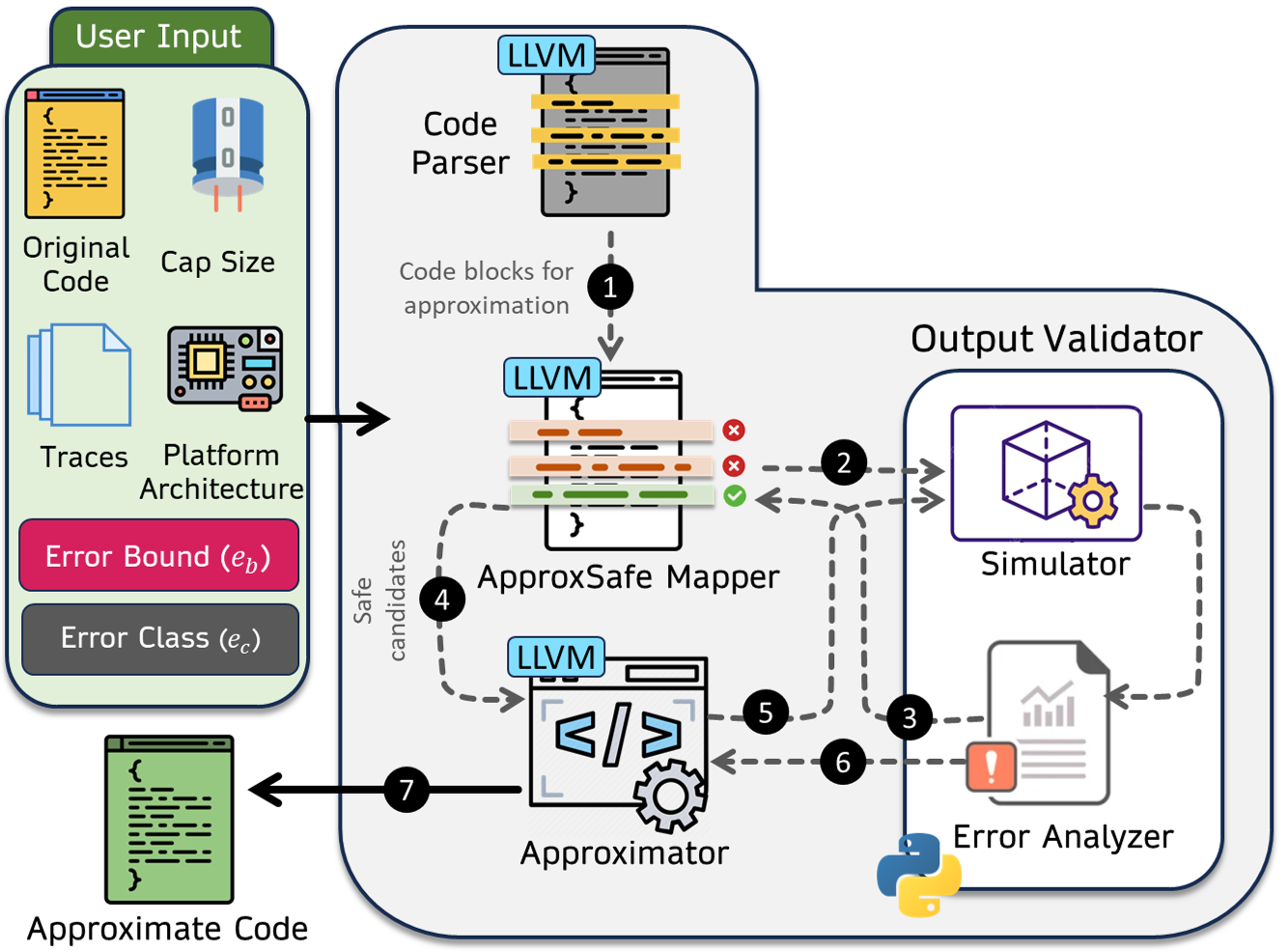}
    \caption{\textit{\tool} architecture.}
    \label{fig:arch}
        \vspace{-0.2in}
\end{figure}

\subsection{System Architecture}

\tool collects six essential inputs from the user: \textit{original application source code} for identifying approximation opportunities, \textit{input trace} or input characteristics to simulate possible application inputs, \textit{error class ($e_c$)} defining acceptable output deviations, \textit{error bound ($e_b$)} for managing the impact of approximations on the accuracy, \textit{platform architecture} (e.g., ARM Cortex M or MSP430) for precise energy consumption estimation, and \textit{capacitor size} to determine the minimum energy required to prevent non-progressive states. These inputs enable \tool to finely balance accuracy and energy efficiency within the specific requirements of batteryless IoT contexts. 

These inputs enable \tool to finely balance accuracy and energy efficiency within the specific requirements of batteryless IoT contexts. \tool's components, including the \firstcomp, \secondcomp, \thirdcomp, and \fourthcomp, work together to analyze and optimize code for approximations. Each component's role is explained below (see Figure~\ref{fig:arch}).

\subsubsection{\textbf{\firstcomp}}

It identifies basic code blocks, such as variable declarations, function definitions, and loops, preparing them for further analysis. Its main role is to scan the source code, gathering these blocks into a list for further analysis.

\subsubsection{\textbf{\thirdcomp}}

It assesses the performance of approximated applications and their deviation from expected outcomes, using a \textit{simulator} and an \textit{error analyzer}.

\noindent
\textbf{Simulator.} 
Executes the code across input traces to determine the frequency of checkpoints, using emulators like \textit{Renode} and \textit{MSPSim} to calculate cycles and energy required for ARM Cortex M-series and MSP430 platforms, which are predominantly utilized in batteryless systems due to their low current consumption characteristics. The simulator's primary task is to determine the number of checkpoints needed for code execution. This information helps in selecting and optimizing approximation strategies. It incorporates the capacitor size to simulate its charge and discharge cycles, reflecting the energy dynamics essential to the checkpointing system.

\noindent
\textbf{Error analyzer.} 
This component analyzes outputs from the original and approximated programs to calculate the approximation-induced error. It uses the programmer-specified threshold $e_c$ to evaluate error across three key output types for batteryless IoT applications: numeric, text, and image, accommodating diverse use cases like environmental sensing and edge machine learning (cf. Table~\ref{tab:error-classes}).

The error metric, $e_m$, is defined as:

\begin{equation}
e_m = \frac{\lvert e_o - e_a\rvert}{e_o}
\end{equation}

where $e_o$ and $e_a$ are the average errors from the original and approximated programs, respectively. Ranging between 0 and 1, $e_m$ measures the impact of approximation, providing a mechanism to regulate approximation strength via $e_b$. Additionally, a checkpoint reduction ratio, $c$, quantifies execution efficiency comparisons:

\begin{equation}
c = \frac{c_a}{c_o}
\end{equation}

where $c_o$ and $c_a$ are the checkpoints required by the original and approximated programs, respectively. Generally, $c$ values lie between 0 and 1, allowing the \thirdcomp to relay both $e_m$ and $c$ to associated components.

\begin{table}
\vspace{0.06in}
             \centering
             \caption{Output types and error classes of \tool}
             \resizebox{0.28\textwidth}{!}{%
             \begin{tabular}{cc}
                 \toprule
                 Program Output & Error Class \\
                 \midrule
                 \multirow{3}{*} {$\geq 1$ Numeric Values} & Euclidean Distance \\ 
                 & Manhattan Distance \\
                 & Root Mean Square Error\\
                 \midrule
                 \multirow{2}{*}{Text} & Word Error Rate\\
                 & Levenshtein Distance \\
                 \midrule
                 \multirow{3}{*}{Image} & F1-Score \\
                 & Pixel Error Rate \\
                 & Structural Similarity Index \\ 
                 \bottomrule
             \end{tabular}}
             
             \label{tab:error-classes}
                 \vspace{-0.15in}
         \end{table}

\subsubsection{\textbf{\secondcomp}}
\label{approx-safe-mapper}
It identifies blocks of code that are safe for approximations. It evaluates the code blocks identified by the \firstcomp and applies preliminary approximations to assess their viability. Using the \thirdcomp to evaluate the code on user-provided traces, the \secondcomp removes blocks from the search space that: (i) do not improve program efficiency, i.e., do not decrease the checkpoint ratio, $c$; (ii) result in an error, $e_m$, that exceeds the programmer-specified $e_b$; or (iii) cause program crashes, such as infinite loops, memory errors, or illegal address accesses.

\subsubsection{\textbf{\fourthcomp}}
The \fourthcomp refines approximations identified by the \secondcomp, methodically applying and fine-tuning them for optimal performance. It uses the \thirdcomp to assess the impact of various approximation techniques and parameters on the system. With a curated list of code blocks deemed safe for approximation, the \fourthcomp adjusts approximation intensity iteratively across all safe blocks until the approximation-induced error, $e_m$, exceeds the programmer-defined threshold, $e_b$, initiating a reset and evaluation of a new approximation set.

Initial approximations cause minimal errors, $e_m$, but as intensity increases, predictability declines, leading to significant errors or potential program crashes. Conversely, the checkpoint reduction ratio, $c$, decreases at a diminishing rate due to the law of diminishing returns, where benefits of further approximations decrease. Both quantities, modeled as functions of approximation intensity, are depicted in Equation \ref{eq:error} and Equation \ref{eq:performance}, where $e$ is the Euler's number and $a$ is the intensity of approximations. 

\begin{equation}
e_m = e^{\frac{a}{2}} -1
\label{eq:error}
\end{equation}

\begin{equation}
c = e^{-2a}
\label{eq:performance}
\end{equation}


\begin{algorithm}
\caption{Sweeping the approximations search space}\label{approximator}
\scriptsize
\begin{algorithmic}[1]
\Require
\Statex $S: $ list of approximation-safe blocks
\Statex $f: $ source-file
\Statex $L: $ loop perforation techniques
\Statex $M: $ approximate memoization
\Statex $e_b: $ error bound
\Statex $T: $ input trace
\Ensure
\Statex $D: $ A list of approximation techniques, their parameters, and corresponding performance.
\Statex
\State $D \gets \emptyset$
\State $e_t \gets 0$ \Comment{Error for current parameters}
\ForAll{$(l,m) \in L \times M$ }
\State $l_p \gets $ \Call{Init Perforation}{$l$} \Comment{Initialize perforation factor}
\State $f_p \gets $ \Call{Init Memoization}{$m$} \Comment{Initialize memoization factor}
\While{$e_t < e_b$}
\ForAll{loop $ \in S$}
\State $f_a \gets $ \Call{Approximate}{loop, $l_p$}
\EndFor
\ForAll{function $ \in S$}
\State $f_a \gets $ \Call{Approximate}{function, $f_p$}
\EndFor
\State $e_t,c \gets$ \Call{Output Validator}{$f, f_a, T$}
\State $D \langle l,m,l_p,f_p \rangle \gets e_t,c$
\State $l_p \gets $ \Call{Increment Perforation}{$l_p$}
\State $f_p \gets $ \Call{Increment Memoization}{$f_p$}
\EndWhile
\EndFor
\State \Return $D$
\end{algorithmic}
    
\end{algorithm}

 We leverage these functions, hypothesizing that optimal performance is achieved where error and checkpoint reduction are minimized simultaneously. At this point, we find the best balance between error and performance, maximizing checkpoint savings with minimal deviation from expected output. Echoing the bias-variance trade-off in machine learning, our strategy aims to minimize the combined sum of these functions ($min(e_m + c)$), allowing the \fourthcomp to determine the most effective approximation strategy.

\section{Implementation}
 \tool harnesses LLVM \cite{llvm} for code parsing and compiler passes, alongside a Python framework for simulating applications. This synergy enables effective source code modifications and evaluation of application approximations.

\subsection{Parsing and Modifying the Source Code}
\tool leverages LLVM's abstract syntax tree (AST) to identify essential code blocks. \firstcomp isolates and eliminates blocks irrelevant to \tool's function. Using LLVM compiler passes, \tool modifies the source code for specific approximation techniques. \secondcomp applies these compiler passes, forwarding modifications to \thirdcomp for execution simulation and deviation analysis. Blocks that do not improve the checkpoint reduction ratio ($c$) or exceed the error threshold ($e_b$) are excluded from further approximation. \secondcomp compiles a list of viable blocks for \fourthcomp's consideration.\fourthcomp utilizes custom LLVM passes for approximation, monitoring each technique's error ($e$) and adjusting strategies as necessary to find the most effective techniques and parameters.

\subsection{Performance and Accuracy Evaluation}
\tool uses \texttt{Renode}~\cite{renode} and \texttt{MSPSim}~\cite{mspsim}, cycle-accurate emulators, to evaluate approximated code across various microcontroller architectures. \texttt{Renode} supports ARMv7, ARMv8, x86, RISC-V, SPARC, POWER, and XTENSA architectures, while \texttt{MSPSim} focuses on MSP430. These emulators offer a reliable basis for performance analysis, ensuring generalizability across a wide range of batteryless IoT platforms.

\tool integrates a checkpointing simulator, based on \texttt{MementOS}~\cite{mementos}, to mimic energy consumption dynamics using a user-supplied voltage trace and capacitor sizes. The simulator obeys the basic capacitor equations for charging and discharging. This simulator predicts the necessary checkpoints for application completion. The choice of the checkpointing system is irrelevant, as \tool aims to reduce computational cycles, enhancing the energy efficiency of any checkpointing system. Using \texttt{Renode} and \texttt{MSPSim}, \tool efficiently manages application emulation and output tracking. Error classes outlined in Table \ref{tab:error-classes} facilitate detailed comparisons between original and approximated code outcomes, enhancing accuracy assessment.

This integrated framework, combining LLVM's capabilities with \texttt{Renode} and \texttt{MSPSim}'s simulation tools, ensures comprehensive and automated approximation evaluation within \tool.

\section{Evaluation}
\label{sec:eval}

The evaluation of \tool is conducted in two phases to thoroughly assess its functionality. Initially, we analyze the effectiveness of its approximation strategies in reducing checkpoint frequency across two applications relevant to batteryless IoT systems.
Next, we evaluate \tool's accuracy in predicting checkpoint needs under simulated real-world energy scenarios using a specialized testbed, ensuring its applicability in varying power conditions.




\subsection{Performance Benchmarks}
\label{sec:Performance_Benchmarks}
We now report on \tool's performance across the two benchmark applications, each evaluated with four different capacitor sizes. The starting capacitor size varies for each application, determined by the smallest size that supports the application's computational load, reflecting the unique energy requirements of each. These evaluations were conducted using the STM32 L152RE board \cite{nucleol152re}, which utilizes the Cortex-M architecture, serving as our target platform.

\begin{figure}
     \centering
     \begin{subfigure}[b]{0.22\textwidth}
         \centering
         \includegraphics[width=\textwidth]{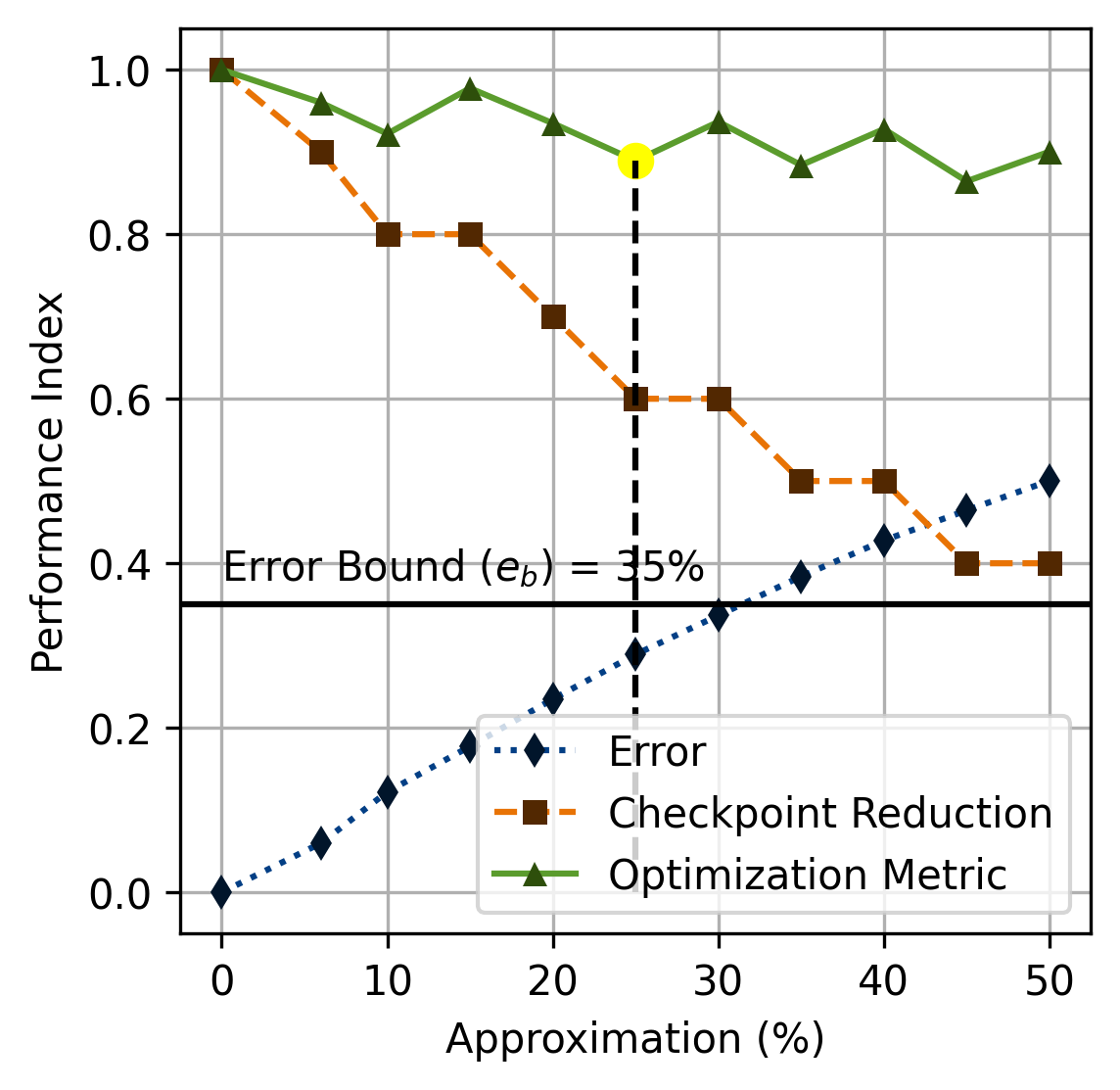}
         \caption{SUSAN - $220\mu F$}
         \label{fig:susan}
     \end{subfigure}
     \begin{subfigure}[b]{0.22\textwidth}
         \centering
         \includegraphics[width=\textwidth]{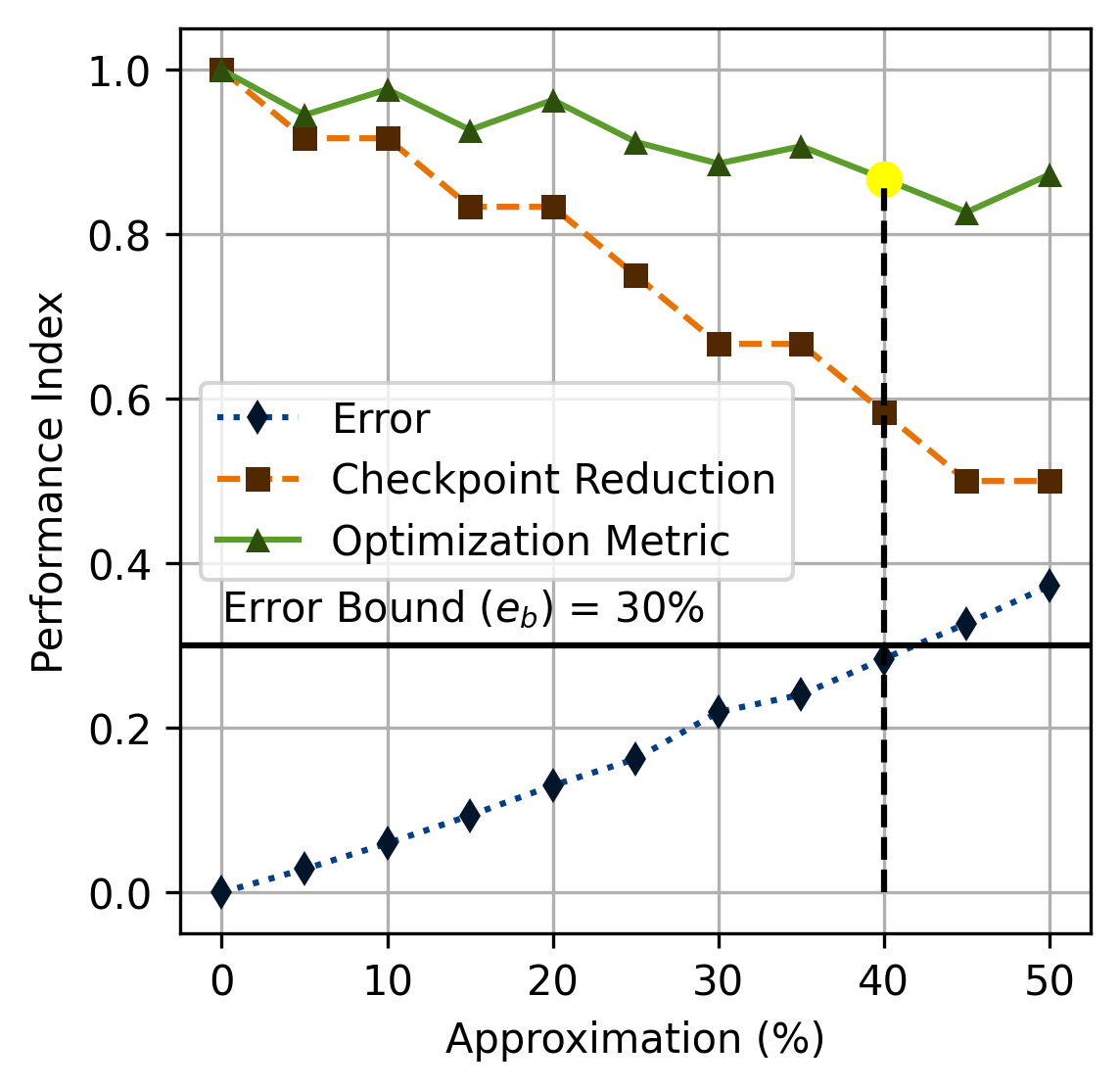}
         \caption{LQI - $220 \mu F$}
         \label{fig:lqi}
     \end{subfigure}
    
        \caption{Performance index vs. approximation percentage for two applications with smallest capacitors.}
        \label{fig:performance}
            \vspace{-0.2in}

\end{figure}



\subsubsection{\textbf{SUSAN}}
The Smallest Univalue Segment Assimilating Nucleus (SUSAN) algorithm, widely used in IoT machine learning applications such as facial recognition, object tracking, and 3D mapping, is evaluated using the BSDS500~\cite{MartinFTM01}, a standard benchmark for edge detection. Implemented from the MiBench2 embedded benchmark suite \cite{MiBench}, the SUSAN algorithm is tested with \tool to autonomously determine appropriate blocks and parameters for approximation. With an error bound ($e_b$) of $35\%$ and using the Structural Similarity Index (SSIM) to evaluate image quality, our findings indicate that a capacitor size below $220 \mu F$ fails to support successful checkpointing, establishing this as the minimal viable capacitor size. Simulations are conducted with four standard capacitor sizes starting from $220 \mu F$.

\tool pinpoints two loops within the SUSAN algorithm’s search area as prime approximation candidates. Initial simulations with the smallest capacitor ($220 \mu F$) are illustrated in Figure \ref{fig:susan}, showing the trade-off between approximation factor, checkpoint reduction, and optimization metric. The yellow point in the figure indicates the optimal point. The outcome reveals a reduction in checkpoint frequency from 10 to 6 for $220 \mu F$, from 3 to 2 for $330 \mu F$, from 2 to 1 for $470 \mu F$, and from 1 to 0 for $680 \mu F$, corresponding to reductions of $40\%, 33\%, 50\%, 100\%$ respectively, with SSIM deviations of $28\%, 23\%, 23\%, 6\%$.


\subsubsection{\textbf{Link Quality Indicator (LQI)}}
The LQI algorithm, integral to wireless communications, evaluates the strength and reliability of communication links within embedded wireless networks. It uses a weighted packet delivery ratio to measure the proportion of successfully delivered packets versus total transmitted, offering critical insights for routing, resource allocation, and network optimization.

For testing \tool with the LQI algorithm, we employ a simulated network trace between nodes of varying link strengths, analyzing over $1000$ data points per link with timestamp-based weighting—recent packets are weighted more due to their greater relevance to current link quality. 

\tool autonomously identifies approximation-safe blocks, applying suitable techniques for optimal performance. We input network data, set a $30\%$ error boundary (Root Mean Square Error or RMSE), and find that \tool targets a single block managing packet history and weights. Omitting some iterations, especially those involving older packets, does not significantly impact overall link quality. Initial tests determine $220 \mu F$ as the minimal capacitor size necessary. Through iterative testing, \tool recommends a $35\%$ approximation factor for the smallest capacitor, maintaining an error within the set boundary of $28\%$ and achieving a $40\%$ reduction in checkpoints. Figure \ref{fig:lqi} illustrates \tool's performance on the LQI application. Further evaluations across various capacitor sizes consistently yield the most effective approximation parameters.

\subsection{Testbed Validation: Emulating Real-World}
To validate \tool's simulated performance predictions, we employ a testbed that replicates real-world energy harvesting conditions, crucial for confirming \tool's reliability in diverse energy environments and to ensure repeatability.

\subsubsection{Testbed Configuration}

\begin{figure}
     \centering
         \includegraphics[width=\columnwidth]{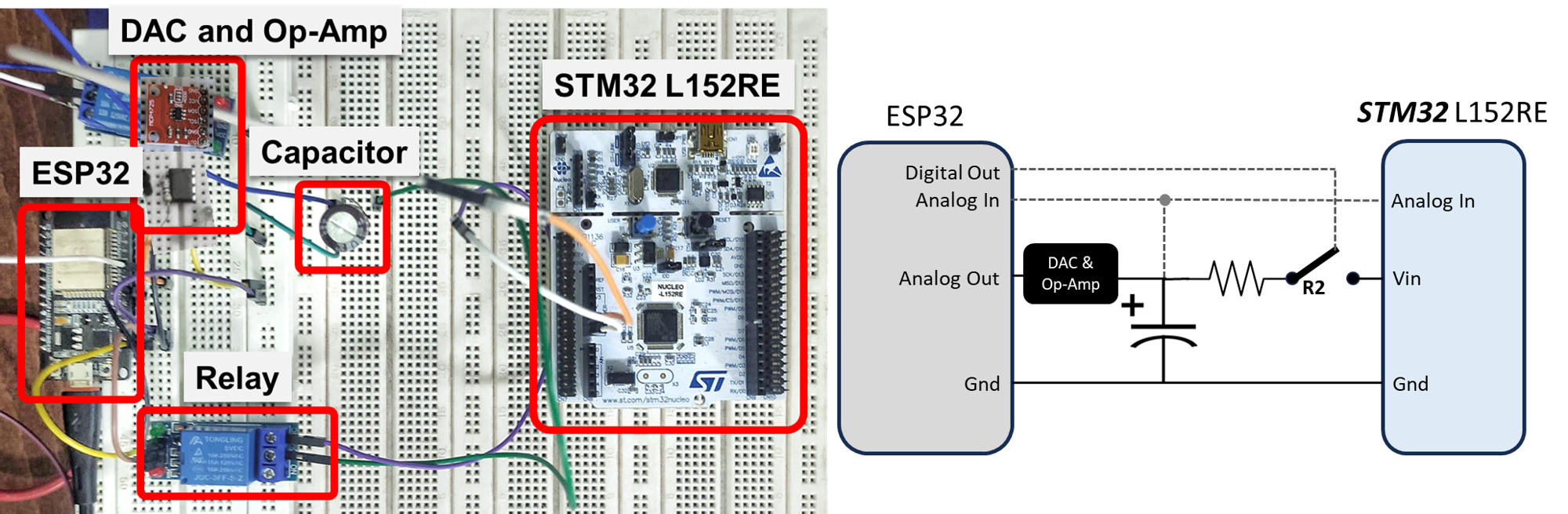}
        \caption{Testbed and schematic for checkpointing validation}
        \label{fig:testbed}
            \vspace{-0.15in}

\end{figure}

Our testbed comprises two primary components: the ESP32 microcontroller, which acts as the control unit, and the STM32 L152RE Nucleo board, which serves as the test device. We have implemented a checkpointing system on the STM32 L152RE to replicate the functionality of \tool's simulator, enabling precise comparative analysis as shown in Figure~\ref{fig:testbed}.

The ESP32 manages the energy supply, delivering an adjustable voltage ranging from 0V to 5V. This is facilitated by an external Digital-to-Analog Converter (DAC) interfaced with the ESP32, paired with an operational amplifier circuit designed to boost the DAC's output to the necessary 0-5V range. This setup effectively simulates a broad spectrum of energy harvesting conditions, allowing precise adjustments to mirror the typical energy constraints encountered. We set the current limit at 6mA using a 500 Ohm resistor to emulate low-power energy harvesting scenarios, ensuring that the supply current reflects a constrained energy environment. The STM32 L152RE board, which draws about 12mA, relies on checkpointing for operation, as it cannot sustain continuous function without it, even when the energy trace is constant.

Energy dynamics are controlled via a relay system that is controlled by ESP32 and  connects the STM32 L152RE board to a capacitor. This system initiates the discharge of the capacitor when its voltage reaches the threshold of 3.3V. When the voltage drops to 1.8V, the relay disconnects, allowing the capacitor to accumulate energy. This mechanism closely emulates real-world scenarios where capacitors charge and discharge concurrently.

\begin{figure}
    \centering
    \includegraphics[width=2in]{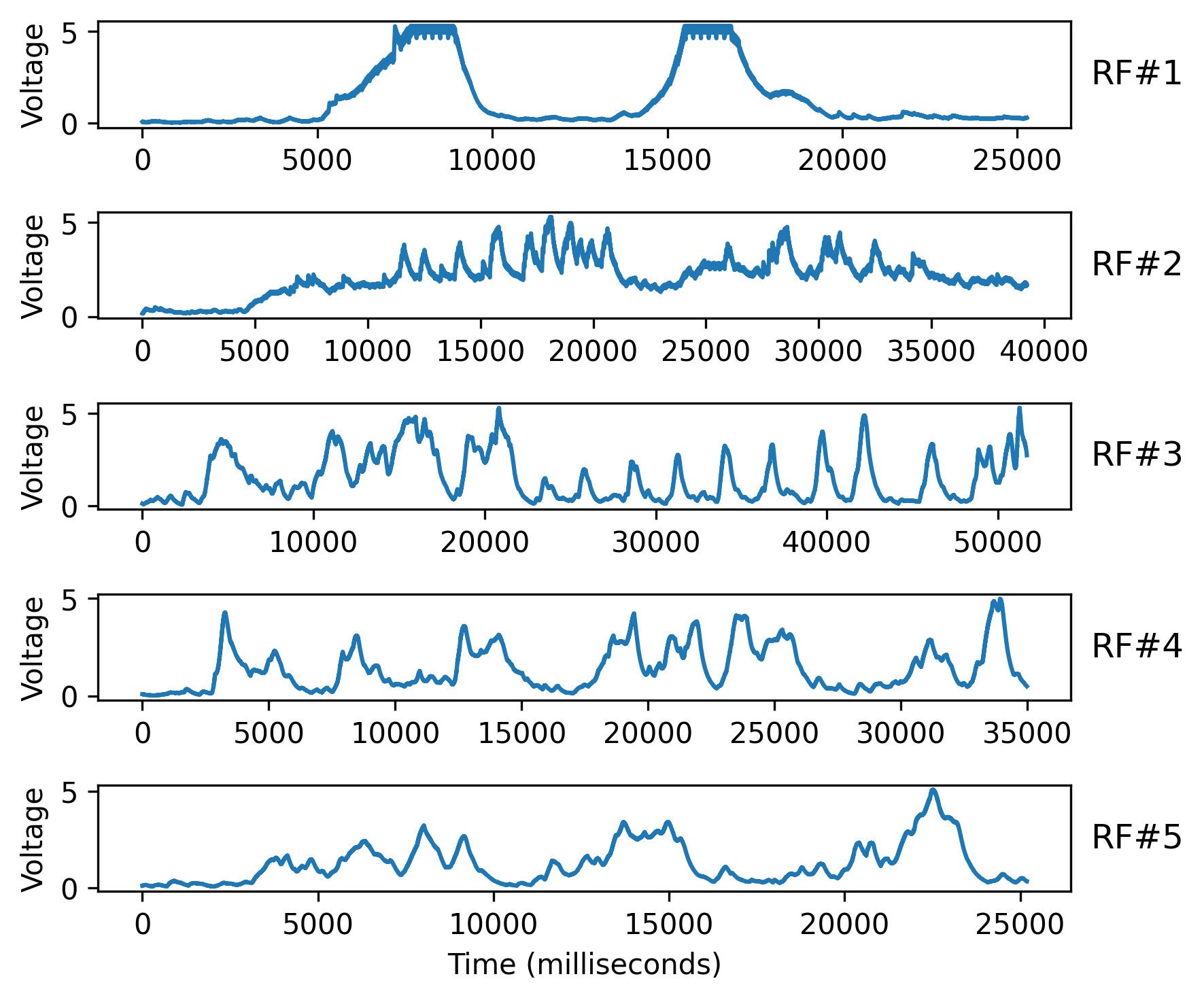}
    \caption{Energy traces}
    \label{fig:traces}
    \vspace{-0.2in}
\end{figure}

To simulate a variety of energy harvesting environments, we utilized five distinct radio frequency (RF) energy traces, originally collected by MementOS and reproduced on our testbed (Figure~\ref{fig:traces}). Throughout the experiment, the ESP32 continuously measures the capacitor voltage at 1 ms intervals to orchestrate the state of the relay, effectively simulating charge and discharge phases. Simultaneously, the STM32 L152RE board also gauges the capacitor voltage, allowing \texttt{MementOS} to determine and trigger the checkpoints dynamically.

\begin{figure*}
     \centering
    \includegraphics[width=6.2in]{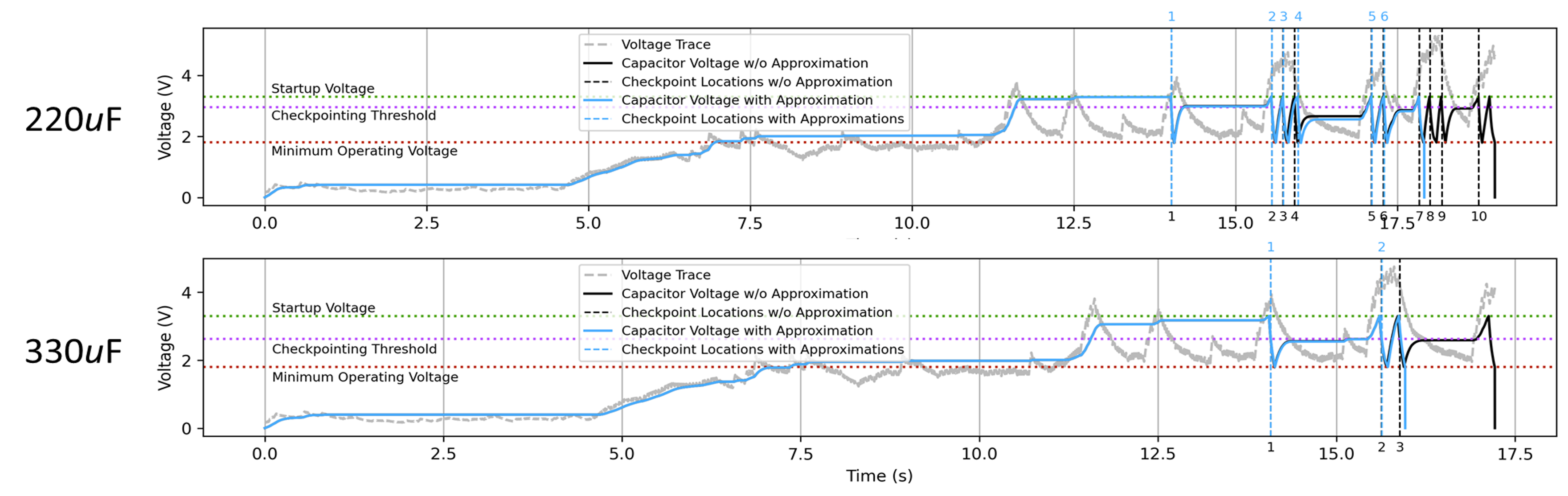}
     
        \caption{Performance of the SUSAN under dynamic energy harvesting with 220uF and 330uF capacitors. The graphs show RF energy harvested voltage and checkpointing instances with and without approximation, validating \tool's efficieny.}
        \label{fig:variable_V_results}
            \vspace{-0.1in}

\end{figure*}

\subsubsection{Results}

We anticipate that the outcomes will corroborate \tool's simulated results, given that the incoming energy's magnitude is significantly lower than the consumption rate, thereby not affecting the capacitor's discharge cycle. The number of checkpoints predicted by our simulator is exactly the same as observed on the testbed, which not only verifies the output of \tool but also demonstrates its effectiveness in reducing the number of checkpoints by reducing computational overhead. This reduction in computational overhead could potentially lead to improved network performance, such as lower latency and increased throughput in data processing tasks.  Figure~\ref{fig:variable_V_results} showcases the performance of the SUSAN application under RF\#2 energy trace across four capacitor sizes, while a summary of results from other traces is available in Table~\ref{tab:combined-traces}.

\begin{table}[]
\centering
\caption{Number of checkpoints on multiple energy traces for SUSAN, LQI, and String Search applications. Each cell contains three values: the number of checkpoints on the testbed (\textcolor{black}{black}), predicted by Approxify’s simulator (\textcolor{cyan}{cyan}), and the original without approximations (\textcolor{red}{red}).
}
\label{tab:combined-traces}
\resizebox{0.38\textwidth}{!}{
\begin{tabular}{c|cc|cc|cc}
\toprule
Traces & \multicolumn{2}{c|}{\cellcolor{susanColor}SUSAN} & \multicolumn{2}{c|}{\cellcolor{lqiColor}LQI} & \multicolumn{2}{c}{\cellcolor{ssColor}String Search} \\ \cline{2-7}
     & 220µF & 330µF & 220µF & 330µF & 47µF & 68µF \\ \midrule
RF\#1  & \textcolor{black}{6}/\textcolor{cyan}{6}/\textcolor{red}{10} & \textcolor{black}{2}/\textcolor{cyan}{2}/\textcolor{red}{3} & \textcolor{black}{7}/\textcolor{cyan}{7}/\textcolor{red}{10} & \textcolor{black}{3}/\textcolor{cyan}{3}/\textcolor{red}{4} & \textcolor{black}{8}/\textcolor{cyan}{8}/\textcolor{red}{10} & \textcolor{black}{2}/\textcolor{cyan}{2}/\textcolor{red}{4} \\
RF\#2  & \textcolor{black}{6}/\textcolor{cyan}{6}/\textcolor{red}{10} & \textcolor{black}{2}/\textcolor{cyan}{2}/\textcolor{red}{3} & \textcolor{black}{7}/\textcolor{cyan}{7}/\textcolor{red}{10} & \textcolor{black}{3}/\textcolor{cyan}{3}/\textcolor{red}{4} & \textcolor{black}{8}/\textcolor{cyan}{8}/\textcolor{red}{10} & \textcolor{black}{2}/\textcolor{cyan}{2}/\textcolor{red}{4} \\
RF\#3  & \textcolor{black}{6}/\textcolor{cyan}{6}/\textcolor{red}{10} & \textcolor{black}{2}/\textcolor{cyan}{2}/\textcolor{red}{3} & \textcolor{black}{7}/\textcolor{cyan}{7}/\textcolor{red}{10} & \textcolor{black}{3}/\textcolor{cyan}{3}/\textcolor{red}{4} & \textcolor{black}{9}/\textcolor{cyan}{9}/\textcolor{red}{11} & \textcolor{black}{2}/\textcolor{cyan}{2}/\textcolor{red}{4} \\
RF\#4  & \textcolor{black}{6}/\textcolor{cyan}{6}/\textcolor{red}{10} & \textcolor{black}{2}/\textcolor{cyan}{2}/\textcolor{red}{3} & \textcolor{black}{7}/\textcolor{cyan}{7}/\textcolor{red}{10} & \textcolor{black}{3}/\textcolor{cyan}{3}/\textcolor{red}{4} & \textcolor{black}{8}/\textcolor{cyan}{8}/\textcolor{red}{10} & \textcolor{black}{2}/\textcolor{cyan}{2}/\textcolor{red}{4} \\
RF\#5  & \textcolor{black}{6}/\textcolor{cyan}{6}/\textcolor{red}{10} & \textcolor{black}{2}/\textcolor{cyan}{2}/\textcolor{red}{3} & \textcolor{black}{7}/\textcolor{cyan}{7}/\textcolor{red}{10} & \textcolor{black}{3}/\textcolor{cyan}{3}/\textcolor{red}{4} & \textcolor{black}{8}/\textcolor{cyan}{8}/\textcolor{red}{10} & \textcolor{black}{2}/\textcolor{cyan}{2}/\textcolor{red}{4} \\
\bottomrule
\end{tabular}}
    \vspace{-0.1in}
\end{table}


\section{What's next?}

Our current work combines known techniques into a robust architecture, serving as a foundational demonstration of feasibility for such systems in batteryless IoT environments. While our automated strategies improve efficiency, they may not achieve the same level of checkpoint reduction and accuracy balance as manual fine-tuning by humans.

To advance this, we are exploring the use of Large Language Models (LLMs) to achieve fine-tuning that approaches human-level precision. LLMs, with their advanced contextual and semantic analysis capabilities, can provide more nuanced and effective solutions. LLMs can interpret code beyond mere syntax, identifying subtle opportunities for approximation that maintain core functionality while enhancing energy efficiency. For instance, in image processing, an LLM might reduce the precision of certain calculations during initial filtering stages, minimally impacting final output quality but significantly saving energy. This even allows LLMs to propose alternative algorithms or functions that achieve computational goals with reduced energy consumption. We also recognize that the offline decisions made by Approxify may be conservative. Ambient energy levels during execution might remain high enough to avoid aggressive approximations. To address this, we plan to integrate an online decision policy that adaptively selects approximation factors based on real-time energy trends.


\section{Conclusions}
We have introduced \tool, a tool aimed at optimizing energy efficiency and maintaining computational accuracy in batteryless IoT. Our evaluation demonstrates that \tool effectively reduces energy consumption and minimizes the need for frequent checkpoints, which are critical challenges in intermittently powered communication systems. Approxify holds significant potential to enhance the performance of batteryless IoTn devices, contributing to the broader discourse on the integration of approximate computing in wireless communication technologies. 

\balance

\bibliographystyle{IEEEtran}
{\footnotesize
\bibliography{citations}}

\end{document}